\documentclass[
    letterpaper,
    nospread,
]{jacow}
\makeatletter%
	\ifboolexpr{bool{xetex}}
	 {\renewcommand{\Gin@extensions}{.pdf,%
	                    .pdf,.jpg,.bmp,.pict,.tif,.psd,.mac,.sga,.tga,.gif,%
	                    .eps,.ps,%
	                    }}{}
\makeatother

\ifboolexpr{bool{xetex} or bool{luatex}} % test for XeTeX/LuaTeX
 {}                                      % input encoding is utf8 by default
 {\usepackage[utf8]{inputenc}}           % switch to utf8

\usepackage[USenglish]{babel}

\usepackage[final]{pdfpages}
\usepackage{multirow}
\usepackage{ragged2e}
\usepackage{subcaption}
\usepackage{bm}

\usepackage{todonotes}
\presetkeys{todonotes}{inline}{}
\ifboolexpr{bool{jacowbiblatex}}%
 {%
  \addbibresource{TUC1C2.bib}
 }{}
\begin{document}

\title{The impact of high-dimensional phase space correlations on the beam dynamics in a linear accelerator%
\thanks{
    This manuscript has been authored by UT- Battelle, LLC under Contract No. DE-AC05-00OR22725 with the U.S. DOE. The US government retains and the publisher, by accepting the article for publication, acknowledges that the US government retains a nonexclusive, paid-up, irrevocable, worldwide license to publish or reproduce the published form of this manuscript, or allow others to do so, for US government purposes. DOE will provide public access to these results of federally sponsored research in accordance with the DOE Public Access Plan (http://energy.gov/downloads/doe-public-access-plan). 
}}

\author{A. Hoover\thanks{hooveram@ornl.gov}, K. Ruisard, A. Aleksandrov, A. Zhukov, S. Cousineau, A. Shishlo \\ Oak Ridge National Laboratory, Oak Ridge, TN, USA}
	
\maketitle

\begin{abstract}
Hadron beams develop intensity-dependent transverse-longitudinal correlations within radio-frequency quadrupole (RFQ) accelerating structures. These correlations are only visible in six-dimensional phase space and are destroyed by reconstructions from low-dimensional projections. In this work, we estimate the effect of artificial decorrelation on the beam dynamics in the Spallation Neutron Source (SNS) linac and Beam Test Facility (BTF). We show that the evolution of a realistic initial distribution and its decorrelated twin converge during the early acceleration stages; thus, low-dimensional projections are probably sufficient for detailed predictions in high-power linacs.
\end{abstract}

\section{Introduction}
\label{sec:introduction}

Predicting halo formation in linear hadron accelerators is complicated by incomplete knowledge of the beam's initial distribution in six-dimensional phase space \cite{Allen_2002, Qiang_2002, Groening_2008}. Denoting the positions by $x$, $y$, $z$ and momenta by $p_x$, $p_y$, $p_z$, the distribution function $f(x, p_x, y, p_y, z, p_z)$ is typically reconstructed from the set of orthogonal two-dimensional projections $\{ f(x, p_x), f(y, p_y), f(z, p_z) \}$:
\begin{equation}\label{eq:decorr}
    f(x, p_x, y, p_y, z, p_z) = f(x, p_x) f(y, p_y) f(z, p_z).
\end{equation}
Alternatively, if only the covariance matrix is known, it is common to assume an analytic, ellipsoidally symmetric distribution function. It is unclear whether such approximations are sufficient to predict the detailed beam evolution at the halo level in high-power linear accelerators.

The research program at the Spallation Neutron Source (SNS) Beam Test Facility (BTF) aims to predict halo formation over a short distance. The BTF contains a replica of the SNS linac front-end, including a negative hydrogen ion source, electrostatic low-energy beam transport (LEBT), and 402.5 MHz radio-frequency quadrupole (RFQ). The medium-energy beam transport (MEBT) is similar to the SNS but has no rebunching cavities and is followed by a 9.5-cell FODO transport line, where halo is expected to form. A suite of high-dimensional and high-dynamic-range phase space diagnostics are located at both ends of the MEBT \cite{Zhang_2020}.

Previous work at the BTF has focused on reconstructing and visualizing the phase space distribution at the MEBT entrance \cite{Cathey_2018, Ruisard_2020, Ruisard_2021, Aleksandrov_2021, Hoover_2023}. These studies have unveiled high-dimensional, intensity-dependent correlations between phase space coordinates. One way to investigate the origin of these features is to simulate the upstream beam evolution. The distribution in the LEBT cannot be measured in the BTF, but we have access to older measurements of $\{ f(x, p_x), f(y, p_y) \}$ from a similar ion source \cite{Ruisard_2020}. By propagating samples from this uncorrelated, unbunched LEBT distribution through the RFQ and MEBT, we obtain a ``model'' distribution that can be compared to direct measurements. 

We have evidence that the model distribution is realistic --- the RFQ and MEBT simulations capture the essential beam dynamics. It follows that we can use the model distribution to estimate the impact of artificial \textit{decorrelation} (see Eq.~\eqref{eq:decorr}) on the beam evolution. In this paper, we summarize our comparisons of the model and measured distributions. We then estimate the influence of the relevant high-dimensional features on the beam evolution in the BTF and as well as in the SNS linac. We conclude that reconstructions from two-dimensional measurements, as in Eq.~\eqref{eq:decorr}, are probably sufficient for detailed predictions in high-power linacs.

\section{PARMTEQ vs. reality}

We use transverse slits, a dipole-slit energy spectrometer, and a bunch shape monitor (BSM) to reconstruct the phase space distribution in the MEBT, 1.3 meters past the RFQ. Six-dimensional measurements still have quite low resolution ($\approx 10^{6}$ points) and dynamic range ($\approx 10^2$) since their first demonstration (although these numbers are expected to improve in the near future \cite{Kutsaev_2021}). Four- and five-dimensional measurements are high-resolution alternatives that capture most correlations of interest. We refer the reader to Refs.~\cite{Ruisard_2020, Ruisard_2021, Hoover_2023} for details. Ref.~\cite{Ruisard_2020} also contains a detailed description of the PARMTEQ \cite{Crandall_1988} model and LEBT distribution used in our RFQ simulations.

\begin{figure*}
    \centering
    
    \begin{subfigure}[]{\columnwidth}
        \centering
        \includegraphics[width=\columnwidth]{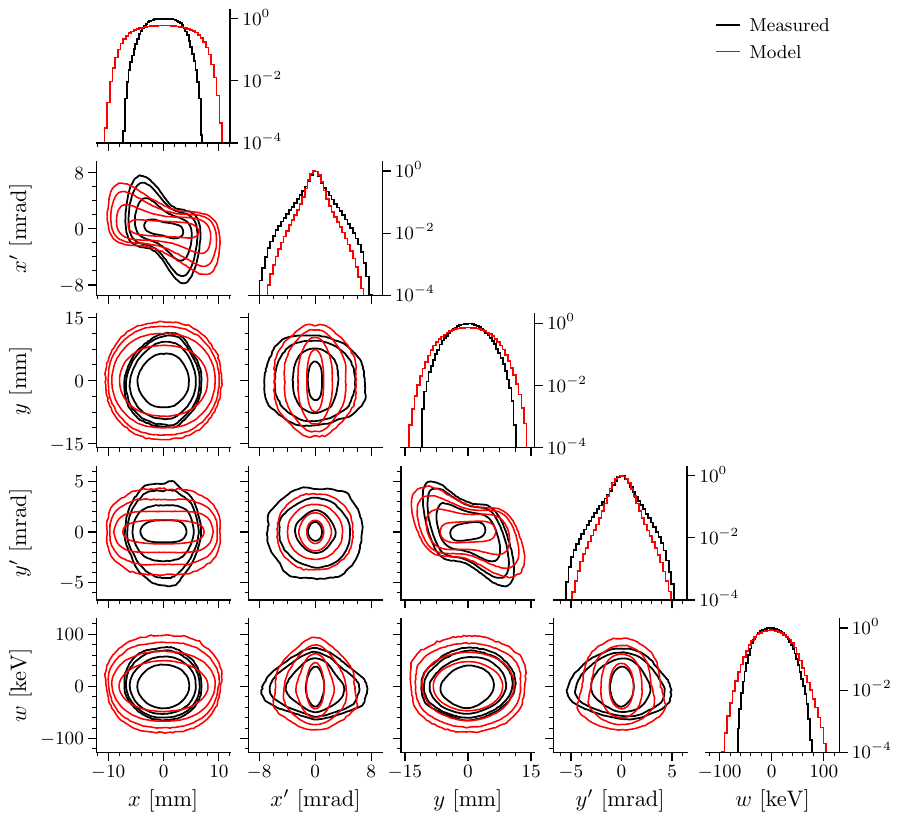}
        \caption{Unnormalized coordinates.}
        \label{fig:compare_5d_corner_a}
    \end{subfigure}
    \hfill
    \begin{subfigure}[]{\columnwidth}
        \centering
        \includegraphics[width=\columnwidth]{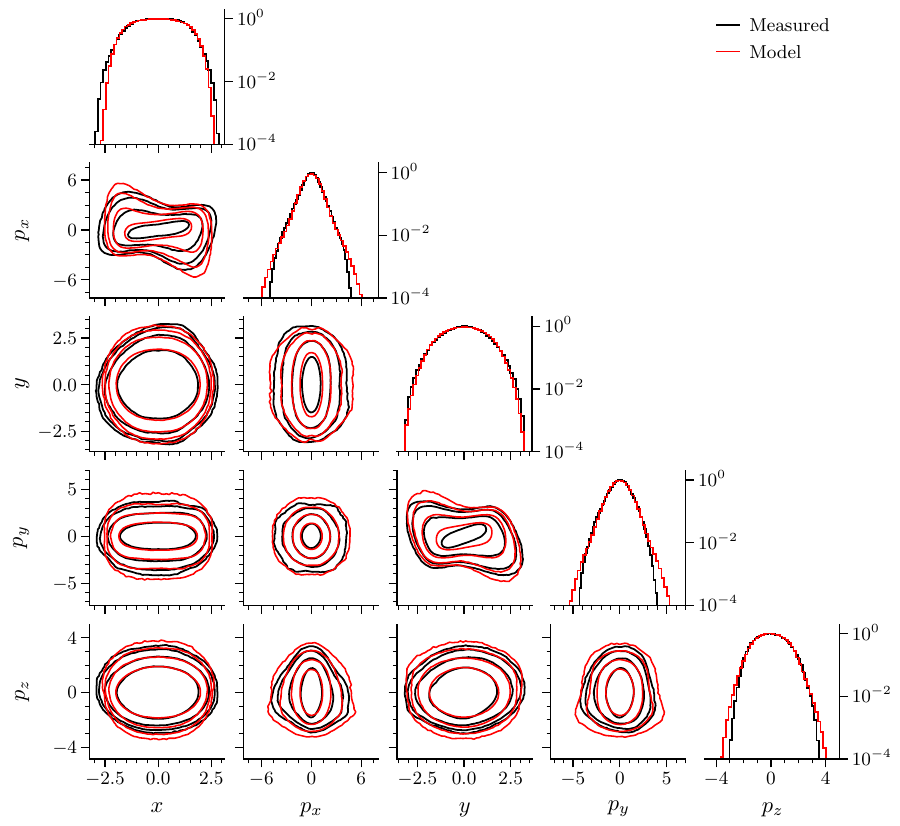}
        \caption{Normalized coordinates.}
        \label{fig:compare_5d_corner_b}
    \end{subfigure}
    \caption{One- and two-dimensional projections of the measured (black) and model (red) five-dimensional phase space distribution. In (b), the coordinates are normalized to unit covariance. The contours vary logarithmically from $10^{-3.0}$ to $10^{-0.5}$ as a fraction of the peak density in each frame.}
    \label{fig:compare_5d_corner}
\end{figure*}

Fig.~\ref{fig:compare_5d_corner} compares the low-dimensional projections of the model distribution to a five-dimensional measurement.
Note that the RFQ model predicts a 42 mA beam current, higher than the 25 mA measured current. Although the predictions are incorrect by a large margin, normalizing both data sets to unit covariance brings the predicted projections fairly close to the measured projections. From now on, we let $x$, $p_x$, $y$, $p_y$, $z$, $p_z$ represent these normalized coordinates.

A striking conclusion drawn from the BTF measurements is that various higher dimensional projections of the six-dimensional phase space distribution are hollow.\footnote{It is well-known from the Kapchinskij-Vladimirskij distribution (a four-dimensional ellipsoidal shell) that a hollow core is easily hidden by low-dimensional projections.} It is, on one hand, unsurprising that inter-plane correlations develop in the RFQ, where a complex bunching process occurs over many focusing periods with strong space charge. On the other hand, some measured features are not entirely intuitive; for example, they contain unexpected asymmetries. Below, we highlight two such features and show that they are also present in the model distribution.

\begin{figure}
    \centering
    \includegraphics[width=\columnwidth]{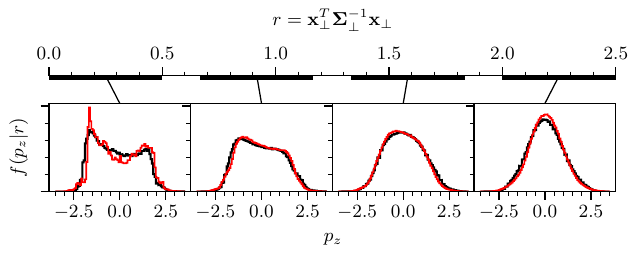}
    \caption{Energy distribution of the measured (black) and model (red) bunch within ellipsoidal shells in transverse phase space. Each shell slice is defined by $r_{min} \le \mathbf{x}^T_{\perp} \mathbf{\Sigma}^{-1}_{\perp} \mathbf{x}_{\perp} \le r_{max}$ for covariance matrix $\mathbf{\Sigma}_{\perp} = \langle \mathbf{x}_{\perp}\mathbf{x}^T_{\perp} \rangle$ and transverse phase space coordinate vector $\mathbf{x}_{\perp} = [x, p_x, y, p_y]^T$.}
    \label{fig:compare_5d_radial}
\end{figure}

First, the longitudinal momentum distribution is bimodal near the transverse core; i.e., the five-dimensional distribution $f(x, p_x, y, p_y, p_z)$ is hollow. We visualize this feature using radial/ellipsoidal shell slices \cite{Hoover_2023} in Fig.~\ref{fig:compare_5d_radial}.
The model reproduces the transition from unimodal to bimodal $p_z$ when moving from the periphery to the core of the four-dimensional transverse distribution. Ruisard et al. \cite{Ruisard_2020, Ruisard_2021} showed that longitudinal hollowing occurs during the free expansion of a six-dimensional Gaussian distribution, which relaxes to a uniform and eventually hollow spatial density. However, they also concluded that this mechanism is not excited in the MEBT after the RFQ. Additionally, a symmetric, freely expanding Gaussian does not reproduce the measured asymmetric sliced energy distribution. Thus, longitudinal hollowing must occur in the RFQ. Initial investigations suggest the hollowing occurs during the initial bunch formation; a more thorough investigation may be the subject of future work.

\begin{figure}
    \centering
    \includegraphics[width=\columnwidth]{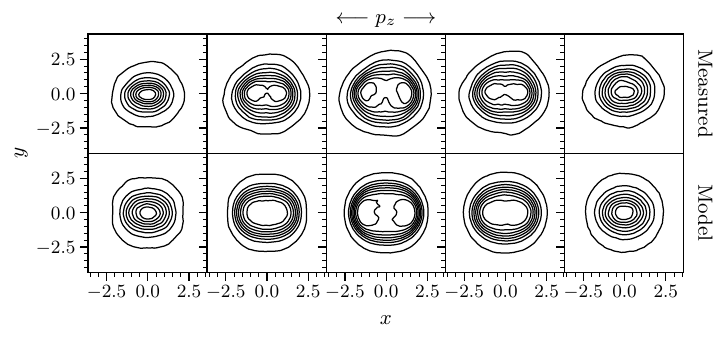}
    \caption{The $x$-$y$ distribution within $p_z$ slices in normalized coordinates. The contours vary linearly from $10^{-2}$ to $10^{0}$ as a fraction of the peak density in each frame.}
    \label{fig:compare_5d_xyz}
\end{figure}

Second, the transverse charge distribution $f(x, y)$ is bimodal when $p_z = 0$; i.e., the three-dimensional distribution $f(x, y, p_z)$ is hollow. Due to the strong $z$-$p_z$ correlation, the charge distribution $f(x, y, z)$ has a similar shape. Again, this feature is present in the model distribution --- see Fig.~\ref{fig:compare_5d_xyz}. 
The transverse hollowing is due to space-charge-driven relaxation of the peaked charge distribution after the RFQ to a uniform and eventually slightly hollow charge distribution at the measurement plane; the hollowing does not require any cross-plane correlations at the MEBT entrance \cite{Hoover_2023}.

We conclude that the RFQ model generates a realistic position-momentum distribution. Thus, we can use the model distribution to investigate the influence of high-dimensional phase space correlations on the beam dynamics. The next section describes two such studies in the SNS linac and the new BTF straight layout.\footnote{The BTF previously used a bent layout; a new straight layout design will be commissioned in late 2023.}

\section{Artificial decorrelation}

We decorrelate an $N$-particle bunch by permuting the particle indices in the $x$-$p_x$, $y$-$p_y$, and $z$-$p_z$ planes:
\begin{equation}
\begin{aligned}
    \{ x_i, {p_x}_i \} &\rightarrow \{ x_i, {p_x}_i \}, \\
    \{ y_i, {p_y}_i \} &\rightarrow \{ y_j, {p_y}_j \}, \\
    \{ z_i, {p_z}_i \} &\rightarrow \{ z_k, {p_z}_k \},
\end{aligned}
\end{equation}
where $i$ is an integer running from $1$ to $N$, and $j,k$ are permutations of those integers. Decorrelation removes all relationships between planes without changing the projections $\{ f(x, p_x), f(y, p_y), f(z, p_z) \}$. Note that decorrelation also creates artificial ``corners'', inflating the six-dimensional phase space volume.

\begin{figure}
    \centering
    \includegraphics[width=\columnwidth]{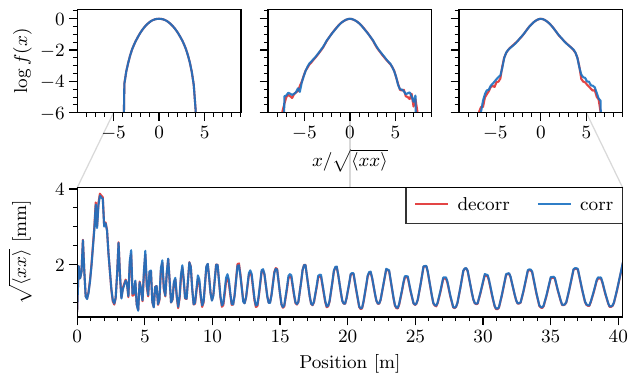}
    \caption{Simulated beam evolution in the SNS linac. Top: horizontal projection $f(x)$ at the beginning, middle, and end of the lattice. The horizontal coordinate is scaled to unit variance. Bottom: rms horizontal beam size as a function of position. The correlated/decorrelated bunch is represented by blue/red lines. (The lines overlap at almost all points.)}
    \label{fig:linac}
\end{figure}

After the bunch is decorrelated, it is straightforward to run two parallel simulations and monitor the divergence between the correlated and decorrelated distributions as they evolve. Here we consider the MEBT and Drift Tube Linac (DTL) sections of the SNS linac. We modelled the linac using PyORBIT \cite{Shishlo_2015} with a transit time factor rf gap model, hard-edged (non-overlapping) quadrupole fields, and FFT space charge solver on a $128 \times 128 \times 128$ mesh. Results are summarized in Fig.~\ref{fig:linac}. 
There are no apparent differences between the correlated and decorrelated beams at the rms level, and the low-dimensional projections are almost identical even in the low-density tails.\footnote{Previous studies produced an intensity-dependent divergence between the correlated/decorrelated horizontal beam sizes by the end of the DTL \cite{Ruisard_2021_ipac}, but these calculations were erroneous. A few particles escaped the stable region of longitudinal phase space, ending up far behind the bunch and eventually lost to a transverse aperture. The grid used in the space charge solver expanded to include these particles, placing almost the entire bunch in one longitudinal bin and generating incorrect space charge forces. The particles escaped at slightly different times in the correlated and decorrelated bunches, leading to an apparent intensity-dependent difference in the rms beam sizes. We fixed this problem by adding a dense array of longitudinal apertures in the MEBT and DTL.} Note that $8 \times 10^6$ simulation particles were tracked, while there are approximately $5 \times 10^8$ particles in each real bunch.

\begin{figure*}
    \centering
    \includegraphics[width=\textwidth]{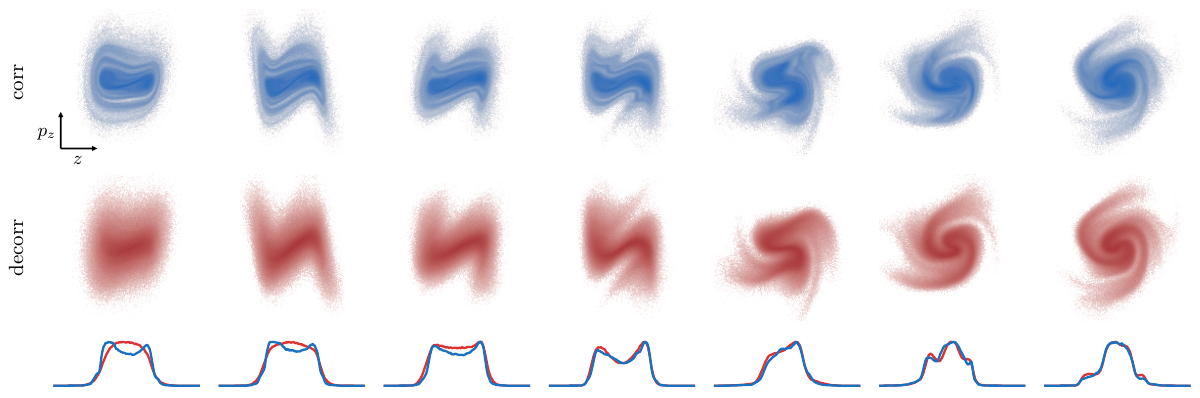}
    \caption{Simulated sliced longitudinal phase space distribution in the first 7 meters of the SNS linac (left to right). The slice selects particles within the rms ellipsoid in transverse phase space. Each distribution is normalized to unit covariance. The scatter plots are shown in logarithmic color scale. One-dimensional projections onto the $z$ axis are overlayed on the bottom row for the initially correlated (blue) and decorrelated (red) bunches.}
    \label{fig:linac_slice}
\end{figure*}

Fig.~\ref{fig:linac_slice} examines the longitudinal phase space distribution near the transverse core during the first 7 meters of transport in the linac.  
In the first half of the transport, the beam is compressed by four rebunching cavities in the MEBT. In the second half, the bunch begins to accelerate in the DTL. The two bunches converge to the same dynamics by the end of the MEBT.

If the lattice and physics models are correct, these results suggest that the measured two-dimensional projections $\{ f(x, p_x), f(y, p_y), f(z, p_z) \}$ are probably sufficient to predict the detailed beam evolution in high-power linear accelerators.

\begin{figure*}[!t]
    \centering
    \includegraphics[width=\textwidth]{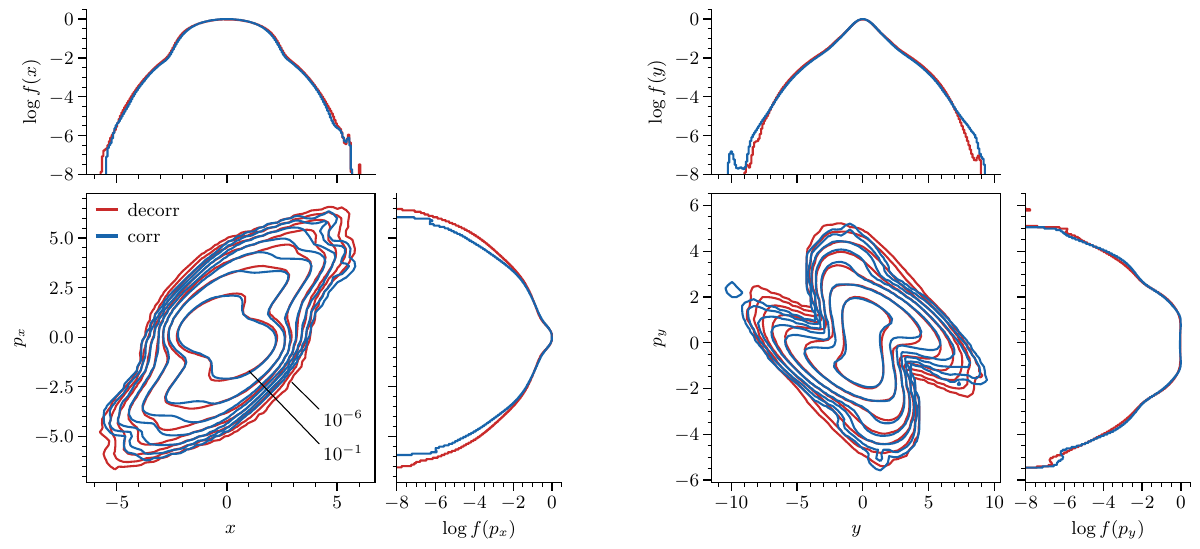}
    \caption{Normalized horizontal and vertical phase space distributions of a simulated bunch at the end of the BTF. The contours of the initially correlated (blue) and decorrelated (red) distributions vary logarithmically.}
    \label{fig:btf}
\end{figure*}

We performed a similar numerical experiment in the BTF lattice, which does not host any rebunching or accelerating cavities. We used the straight-layout design planned for the next experimental run (late 2023). We upsampled the model bunch from $8.5 \times 10^{6}$ to $5.0 \times 10^{8}$ particles, which approaches the real number of particles in a single bunch.\footnote{A variety of upsampling methods were explored, including generative diffusion and normalizing flow models, but in this study, we simply computed and sampled from a six-dimensional histogram with $50^6$ bins. Even with nearly $10^7$ particles, only $0.03\%$ of the histogram bins had nonzero counts; however, adding Gaussian noise to each particle seemed to eliminate most of the unrealistic ``clumping'' in six-dimensional phase space, resulting in a reasonable upsampled bunch.} The optics were set to generate a slightly mismatched envelope in the FODO channel. The simulation ran for 10 hours on four MPI nodes. The $x$-$x'$ and $y$-$y'$ projections at the end of the beamline at displayed in Fig.~\ref{fig:btf}.

The differences between the correlated and decorrelated beams in Fig.~\ref{fig:btf} are small. We emphasize that the phase space structure at the $10^{-6}$ density level is relevant to beam loss in high-power (multi-megawatt) linacs, which must operate below a one-watt-per-meter loss limit, and that this structure can be measured in the BTF \cite{Aleksandrov_2021}.

\section{Discussion}

Our findings in this brief study imply that most of the uncertainty in the initial distribution in the SNS linac (and similar machines) can be eliminated by measuring three orthogonal two-dimensional projections: $\{ f(x, p_x), f(y, p_y), f(z, p_z) \}$. Of course, our tentative conclusions are based on simulations and must be verified with direct measurements. This will be the focus of the upcoming experimental run at the BTF.

These findings also connect to the work in the LEDA experiment, where only the block-diagonal elements of the $6 \times 6$ covariance matrix were measured \cite{Qiang_2002}. In addition to rms-equivalent Waterbag and Gaussian distributions, the authors generated a ``LEBT/RFQ distribution'' analogous to the model distribution defined in this paper. Like us, they found the model distribution's covariance matrix to be significantly different than the measured covariance matrix. Thus, they normalized the model distribution to match the measured covariance matrix. This normalized model distribution generated one-dimensional profiles closer to the measurements but still failed to reproduce the low-density tails, especially when the beam was mismatched. Interestingly, our normalized model distribution agrees quite well with direct high-dimensional phase space measurements. Therefore, it seems that the normalized model distribution in Ref.~\cite{Qiang_2002} should have been quite close to the true distribution. 

A key difference between our studies is that our model distribution was defined by two-dimensional measurements before the RFQ, while the authors of \cite{Qiang_2002} specify that their model distribution was defined at the ion source, perhaps from an analytic distribution function or a simulation of the ion source. This gives motivation to study the sensitivity of the RFQ output bunch to the input distribution; see, e.g., Fig.~11 in Ref.~\cite{Ruisard_2020}.

\bibliographystyle{unsrt}
\bibliography{TUC1C2}

\end{document}